\documentclass[twocolumn]{aastex6}
\usepackage{lineno}
 \usepackage{amsmath}
 \usepackage{amssymb}
 \usepackage{amsthm}
 \usepackage{natbib,aasdefs,url,bm}
 \usepackage{array}
 \usepackage{float}
 \usepackage{graphicx}
 \usepackage{color}
\newcounter{ichi}
\newcounter{ni}
\newcounter{san}
\newcounter{yon}
\newcommand{\vhe}{VHE $\gamma$-rays }
\newcommand{\grb}{GRB 190829A }
\newcommand{\hess}{H.E.S.S. }
\newcommand{\RN}[1]{%
  \textup{\uppercase\expandafter{\romannumeral#1}}%
}

\shorttitle{High-energy gamma-ray emission from GRB 190829A}
\shortauthors{Zhang, Murase, Veres, and M\'esz\'aros}

\begin{document}

\title{External Inverse-Compton Emission from Low-Luminosity Gamma-Ray Bursts: Application to GRB 190829A}

\author{
B. Theodore Zhang\altaffilmark{1,2,3}, 
Kohta Murase\altaffilmark{1,2,3,4},
P\'eter Veres\altaffilmark{5},
P\'eter M\'esz\'aros\altaffilmark{1,2,3}
}
\altaffiltext{1}{Department of Physics, Pennsylvania State University, University Park, Pennsylvania 16802, USA}
\altaffiltext{2}{Department of Astronomy \& Astrophysics, Pennsylvania State University, University Park, Pennsylvania 16802, USA}
\altaffiltext{3}{Center for Multimessenger Astrophysics, Institute for Gravitation and the Cosmos, Pennsylvania State University, University Park, Pennsylvania 16802, USA}
\altaffiltext{4}{Center for Gravitational Physics, Yukawa Institute for Theoretical Physics, Kyoto University, Kyoto, Kyoto 606-8502, Japan}
\altaffiltext{5}{Center for Space Plasma and Aeronomic Research (CSPAR), University of Alabama in Huntsville, Huntsville, AL 35899, USA}

\begin{abstract}
The detection of TeV gamma-ray bursts (GRBs) brought new opportunities for studying the physics of particle acceleration at relativistic shocks.
The \hess telescopes recently observed very-high-energy (VHE) emission from a nearby low-luminosity GRB,
GRB 190829A. 
Follow-up observations with, e.g., \textit{Swift-XRT}, revealed unusual flare activities at $\sim 10^3\rm~s$, which can be caused by a long-lasting central engine.
We show that the VHE emission during the \hess observation time is naturally produced in the external inverse-Compton (EIC) scenario, where seed photons supplied by the flares or other late-time dissipation are upscattered to VHE energies by the non-thermal electrons accelerated at the external forward shock.
Our calculations show that the EIC flare nearly coincides with the late-prompt flare, but extends $\sim3-4$ times longer than the duration of the late-prompt flare.
The preferred kinetic energy and initial Lorentz factor used in our model are $\sim 10^{52}\rm~erg$ and $\sim 20$, respectively.
Understanding the mechanisms of the VHE emission from low-luminosity GRBs will help us constrain the properties of the outflow and the central engine activities, as well as the particle acceleration mechanism.
\end{abstract}

\date{\today}

\section{\label{sec:1}Introduction}
Very high energy (VHE) $\gamma$-rays ($\gamma$-rays with energy greater than $\sim 100\rm~GeV$) have been observed by ground-based Cherenkov telescopes~\citep[e.g.,][]{Hinton:2009zz,Inoue:2013vy}.
The sources of \vhe represent extreme astrophysical accelerators in the universe. In the multimessenger era, the detection of \vhe along with multiwavelength electromagnetic radiation, neutrinos, cosmic rays, as well as gravitational waves can help to unveil the mechanisms of high-energy astrophysical processes~\citep[e.g.,][]{Murase:2019tjj, Meszaros:2019xej}.

The recent detection of two TeV gamma-ray bursts (GRBs), GRB 190114C~\citep{Acciari:2019dxz, Acciari:2019dbx} and GRB 180720B~\citep{Arakawa:2019cfc}, has opened a new window for studying GRBs in the VHE band, providing us with new opportunities to investigate the acceleration and radiation processes~\citep[see][for reviews]{Meszaros:2006rc, Kumar:2014upa}.
\vhe originating from GRBs can be naturally explained with the inverse-Compton (IC) process by relativistic electrons~\citep{Meszaros:1994sd, Dermer:1999eh, Sari:2000zp, Zhang:2001az}. 
The origins of \vhe from GRB 190114C and GRB 180720B have been widely discussed in the framework of the synchrotron self-Compton (SSC) scenario~\citep{Derishev:2019cgi, Arakawa:2019cfc, Fraija:2019whb, Wang:2019zbs, Acciari:2019dbx, Zhang:2019utn, Asano:2020grw}, where the same population of electrons that emit synchrotron photons can also upscatter these photons to much higher energies by a factor of $\sim \gamma_e^2$, where $\gamma_e$ is the electron Lorentz factor~\citep{Blumenthal:1970gc, Rybicki:1986}.

Recently, the High Energy Stereoscopic System (H.E.S.S.) reported the detection of VHE $\gamma$-rays from GRB 190829A with a confidence level of $> 5 \sigma$ ~\citep{2019ATel13052....1D}. The measured redshift of GRB 190829A is $z = 0.0785 \pm 0.005$~\citep{Valeev2019GCN}, which is the nearest GRB detected in the VHE band.
GRB 190829A has been discussed as a possible low-luminosity GRB (LL GRB), which has an isotropic-equivalent luminosity of $L_{\rm iso} \sim 10^{49}\rm~erg~s^{-1}$~\citep{Chand:2020wqt}.
The late-time optical observation revealed an associated broad-line type-\RN{1}c supernovae, SN 2019oyw~\citep{Hu:2020xhu}.
Even though the physical origin of LL GRBs is still controversial, there is a consensus that LL GRBs are related to the deaths of massive stars and have properties between classical GRBs and normal supernovae~\citep{Kumar:2014upa}.
The observed X-ray light curve indicates an unusual long-lasting flare during the time interval $\sim 10^{2} - 10^{4}~\rm s$, which may originate from a late-time central-engine activity ~\citep{Chand:2020wqt}. 

It has been proposed that \vhe can also be produced via the external inverse-Compton (EIC) process in the framework of GRBs, where the seed photons (produced in a region different from the acceleration zone) can be X-ray flares~\citep{Wang:2006eq,He:2011aa}, late-time long-lasting emission~\citep{Murase:2010fq, Murase:2017snw,Veres+12magnetic}, prompt emission~\citep{Murase:2009su}, shock breakout emission~\citep{Wang:2006rp}, hypernova envelope emission~\citep{He:2009rx}, and cocoon emission~\citep{Toma:2009mw,Kimura:2019fae}. 
If the long-lasting X-ray flare photons in \grb catch up with non-thermal electrons accelerated at the external forward shock, then these lower-energy photons can be upscattered to the VHE band.
In the case of anisotropic scattering with target photons impinging from behind, the EIC flux of \vhe is reduced by a factor of two or three compared to the simplest estimate by the product of the scattering optical depth and seed photon flux~\citep{Brunetti:2000wv, Fan:2008zza,Murase:2010fq}. However, if the target photons have a larger energy density compared to synchrotron photons as in the case of late-time flares, then \vhe from the EIC emission can be dominant.

In this work, we study the origin of \vhe in the EIC scenario, with an application to GRB 190829A. The paper is organized as follows.
In Sec.~\ref{sec:general}, we analytically discuss the origin of high-energy emission considering both the SSC and EIC scenarios in the framework of GRB 190829A. In Sec.~\ref{sec:grb}, we show the numerical results and make a comparison to observations. We discuss the implications of this work and give a summary in Sec.~\ref{sec:summary}. Throughout the paper, we use cgs units, and adopt
notations such as $Q_x \equiv Q/10^x$. We use E represents observed photon energy, $\varepsilon = E (1+z)$ is the photon energy measured in the cosmic rest frame, and $\varepsilon^\prime = \varepsilon / \Gamma$ is the photon energy measured in the source comoving frame which have Lorentz factor $\Gamma$.

\section{High-energy afterglow emission from GRBs}\label{sec:general}
We consider a relativistic outflow that has isotropic-equivalent energy $\mathcal{E}_k$ and initial Lorentz factor $\Gamma_0$, propagating into an external medium with constant density $n_{\rm ex}$.
In this work, we numerically calculate afterglow dynamics in detail for both relativistic and non-relativistic regimes. However, in the following text, we mainly discuss the results using the self-similar evolution phase of the adiabatic blast wave~\citep{Blandford:1976uq} to explain essential results. We numerically calculate EIC emission following \citet{Murase:2009su} and \citet{Murase:2010fq}, and we perform numerical calculations for synchrotron emission as well. 

The radius of the forward shock is estimated to be
\begin{equation}
R \simeq 2.2 \times 10^{17}~\mathcal{E}_{k, 52}^{1/4} n_{\rm ex}^{-1/4} t_{z, 4}^{1/4}\rm~cm,
\end{equation}
where $t_{z, 4} = t_4 / (1 + z)$ is the redshift-corrected observation time and the bulk Lorentz factor is
\begin{equation}
\Gamma \simeq 14~\mathcal{E}_{k, 52}^{1/8} n_{\rm ex}^{-1/8} t_{z, 4}^{-3/8},
\end{equation}
where the relation between radius $R$ and observation time $t$ is $R\approx 4\Gamma^2 c t_z$ which takes into account the history of the blast wave evolution and the contribution from a range of angles within $1/\Gamma$ cone~\citep{Waxman:1997yv, Panaitescu:1997jf, Sari:1997qe}.

In the external forward shock model, a fraction of thermal electrons can be injected into the acceleration process and is able to accelerate to higher energies via the diffusive shock acceleration (DSA) mechanism~\citep[e.g.,][]{Drury:1983zz, Blandford:1987pw}.
The lower-energy synchrotron photons and external photons can then be upscattered to the VHE band by high-energy non-thermal electrons.
The luminosity of the newly accelerated electrons can be estimated to be $L_e \sim \epsilon_e 4\pi R^2 c \Gamma^2 U_{\rm sh}^\prime \sim 2 \times 10^{47} \epsilon_{e, -1} \mathcal{E}_{k, 52} t_{z, 4}^{-1}\rm~erg~s^{-1}$, where $\epsilon_e$ is the fraction of internal energy that goes into shocked electrons and $U_{\rm sh}^\prime \sim 2\Gamma^2 n_{\rm ex} m_p c^2$ is the comoving internal energy density~\citep{Wang:2006eq, Fan:2008zza}.
We approximate the synchrotron luminosity to be
\begin{equation}
L_{\rm syn} \sim \frac{1}{1+Y_{\rm tot}} \zeta_\gamma L_e,
\end{equation}
where $\zeta_\gamma$ is the fraction of electron energy that is radiated, $Y_{\rm tot} = Y_{\rm SSC} + Y_{\rm EIC}$ is the total Compton parameter, $Y_{\rm SSC}$ is the SSC Compton parameter defined as the ratio between the SSC luminosity and the synchrotron luminosity, and $Y_{\rm EIC}$ is the EIC Compton parameter defined as the ratio between the EIC luminosity and the synchrotron luminosity. 
In the fast-cooling regime, the value of $\zeta_\gamma$ equals to 1, while in the slow cooling regime $\zeta_\gamma \sim (\gamma_{e,c} / \gamma_{e,m})^{2-s}$ where $s$ is the electron spectral index~\citep{Sari:2000zp}. 
The minimum electron Lorentz factor is
\begin{eqnarray}
\gamma_{e,m} &\approx& (\epsilon_e/f_e) g(s) (m_p/m_e) (\Gamma - 1) \nonumber \\ &\sim& 400\mathcal{E}_{k,52}^{1/8} n_{\rm ex}^{-1/8}  \epsilon_{e, -1} f_{e}^{-1} t_{z, 4}^{-3/8},
\end{eqnarray}
where $\epsilon_e$ is the energy fraction of internal energy that goes into electrons and $f_e$ is the number fraction of electrons that are accelerated, and $g(s) = (s-2)/(s-1)$ for $s > 2$ (we adopt $s = 2.2$).
The maximum electron Lorentz factor is limited by the cooling process,
\begin{eqnarray}
\gamma_{e,M} &\approx& (6\pi e / (\sigma_T B \eta (1 + Y_{\rm tot})))^{1/2} \nonumber \\ &\sim& 9 \times 10^8 \eta^{-1/2} (1+Y_{\rm tot})^{-1/2} \mathcal{E}_{k,52}^{-1/16} n_{\rm ex}^{-3/16} \epsilon_{B,-5}^{-1/4} t_{z, 4}^{3/16},
\end{eqnarray}
where $\eta$ is the acceleration efficiency, which depends on details the acceleration mechanism~\citep[see, e.g., Equation~14 of ][]{Asano:2020grw}, and $\epsilon_B$ is the energy fraction of internal energy that is converted into the magnetic energy.
The electron cooling Lorentz factor is,
\begin{eqnarray}
\gamma_{e,c} &\approx& \frac{6\pi m_e c}{ (1 + Y_{\rm tot})  \sigma_T \Gamma B^2  t_z} \nonumber \\ &\sim& 2 \times 10^7 (1+Y_{\rm tot})^{-1} \mathcal{E}_{k,52}^{-3/8} n_{\rm ex}^{-5/8} \epsilon_{B,-5}^{-1} t_{z, 4}^{1/8}.
\end{eqnarray} 
The non-thermal electrons are in the fast cooling regime for $\gamma_{e,m}>\gamma_{e,c}$, where nearly all of the injected electrons cool during the dynamical time. When $\gamma_{e,m}<\gamma_{e,c}$, electrons are in the slow cooling regime.
Note in the following analytical estimates, we mainly consider the slow cooling regime, while our numerical code can treat both slow- and fast-cooling cases self-consistently.
Note that the fraction of electron energy that was radiated away in the slow cooling case is estimated to be $\zeta_\gamma \sim 0.1 (1 + Y_{\rm tot})^{s-2} \mathcal{E}_{k,52}^{-\frac{2-s}{2}} n_{\rm ex}^{-\frac{2-s}{2}} \epsilon_{B, -5}^{s-2} \epsilon_{e, -1}^{s-2} f_e^{2-s} t_{z, 4}^{\frac{2-s}{2}}$.
The characteristic energy of the synchrotron emission is estimated to be $E_m \approx [e B / (2 \pi m_e c)] \gamma_m^2 \Gamma / (1 + z) \simeq 4.2 \times 10^{-4}~\mathcal{E}_{k,52}^{1/2} \epsilon_{e, -1}^2 f_{e}^{-2} \epsilon_{B,-5}^{1/2} t_4^{-3/2} (1 + z)^{1/2} \rm~eV$ 
and the corresponding cooling energy is $E_c \approx [e B / (2 \pi m_e c)] \gamma_c^2 \Gamma  / (1 + z) \simeq 10^6~(1+Y_{\rm tot})^{-2} \mathcal{E}_{k,52}^{-1/2} n_{\rm ex}^{-1} \epsilon_{B,-5}^{-3/2} t_4^{-1/2} (1 + z)^{-1/2} \rm~eV$.
The observed X-ray luminosity at the keV band will evolve as $L_X \propto t^{(3-3s)/4}$ for $E_m < E_X < E_c$~\citep{Zhang:2005fa}.

The SSC luminosity can be estimated as
\begin{eqnarray}
L_{\rm SSC} \sim \frac{Y_{\rm SSC}}{1+Y_{\rm tot}} \zeta L_e.
\end{eqnarray}
Note that $\epsilon_e \gg \epsilon_B$ is one of the conditions for dominant SSC emission 
~\citep{Sari:1997qe,Zhang:2001az}.
In the Thomson limit, the value of $Y_{\rm SSC}$ remains constant as a function of $\gamma_e$~\citep{Sari:2000zp}, and the SSC luminosity will evolve as $L_{\rm SSC} \propto \zeta_\gamma L_e \propto t_z^{-s/2}$.
Thus, we can expect that the SSC light curve will have a very similar trend as the X-ray light curve for $s > 2$.
The characteristic energies of the SSC emission can be estimated as $E_m^{\rm SSC} \approx 2 \gamma_{e,m}^2 E_m \simeq 1.4 \times 10^2~\mathcal{E}_{k,52}^{3/4} \epsilon_{e, -1}^4 f_{e}^{-4} \epsilon_{B,-5}^{1/2} t_{4}^{-9/4} (1 + z)^{-5/4} \rm~eV$ and $E_c^{\rm SSC} \approx 2 \gamma_{e,c}^2 E_c \simeq 9.5 \times 10^{20}~(1+Y_{\rm tot})^{-4} \mathcal{E}_{k,52}^{-5/4} n_{\rm ex}^{-9/4} \epsilon_{B,-5}^{-7/2} t_{4}^{-1/4} (1 + z)^{-3/4} \rm~eV$.
In the Thomson limit, the SSC energy spectrum in the slow cooling case is
\begin{equation}\label{spec_SSC}
F_E^{\rm SSC} = F_{E, \rm max}^{\rm SSC}  \begin{cases} (\frac{E}{E_m^{\rm SSC}})^{\frac{1}{3}}, & E <  E_m^{\rm SSC} \\ (\frac{E}{E_m^{\rm SSC}})^{-\frac{(s-1)}{2}}, &  E_m^{\rm SSC} < E < E_c^{\rm SSC} \\ \left(\frac{E_c^{\rm SSC}}{E_m^{\rm SSC}} \right)^{\frac{1-s}{2}} \left(\frac{E}{E_c^{\rm SSC}}\right)^{-\frac{s}{2}}, & E > E_c^{\rm SSC} \end{cases},
\end{equation}
where $F_{E, \rm max}^{\rm SSC} \sim \tau_T F_{E, \rm max}^{\rm syn}$ and $\tau_T \sim (1/3) \sigma_T R n_{\rm ex} f_e \sim 5 \times 10^{-8} \mathcal{E}_{k, 52}^{1/4} n_{\rm ex}^{3/4} f_e t_{z, 4}^{1/4}$ is the electron scattering optical depth~\citep{Sari:2000zp}.
The peak synchrotron flux can be calculated as $F_{E, \rm max}^{\rm syn}\approx (1+z) N_e P_{\varepsilon, \rm max} / 4 \pi d_L^2$ where $P_{\varepsilon, \rm max}\approx P(\gamma_{e,m})/\varepsilon_m = (c\sigma_T/6\pi)\gamma_{e,m}^2 B^2 \Gamma^2 / \varepsilon_m$ is the synchrotron emission power per electron~\citep{Sari:1997qe} and $\varepsilon_m = E_m (1+z)$.

Similarly, the EIC luminosity is
\begin{eqnarray}
L_{\rm EIC} \sim \frac{Y_{\rm EIC}}{1+Y_{\rm tot}} \zeta_\gamma L_e,
\end{eqnarray}
The EIC light curve depends on the time evolution of external photons. 
For prompt or flare photons as targets, the observed EIC light curve usually appears as an extended bump compared to the SSC light curve~\citep{Murase:2009su}. 
On the other hand, for long-lasting photons as target photons, the EIC light curve is flatter than the SSC light curve which may dominate at later times~\citep{Murase:2010fq}.
Assuming that the external photons can be described as a broken power law with break energy $E_b$, then the characteristic energies of EIC emission are $E_m^{\rm EIC} \approx 2 \gamma_{e,m}^2 E_b \simeq 3.3 \times 10^7~\mathcal{E}_{k,52}^{1/4} n_{\rm ex}^{-1/4}  \epsilon_{e, -1}^2 f_{e}^{-2} t_{4}^{-3/4} (1+z)^{3/4} E_{b, 2} \rm~eV$ and $E_c^{\rm EIC} \approx 2 \gamma_{e,c}^2 E_b \simeq 8.6 \times 10^{16}~(1+Y_{\rm tot})^{-1} \mathcal{E}_{k,52}^{-3/4} n_{\rm ex}^{-5/4} \epsilon_{B,-5}^{-2} t_{4}^{1/4} (1+z)^{-1/4} E_{b, 2} \rm~eV$.
We model the energy spectrum of the flare as a broken power law without going into the details of the emission mechanism, 
\begin{equation}\label{eq:fl}
F^{\rm fl}_E 
= F^{\rm fl}_{E_b} (t)
\begin{cases}
\left(\frac{E}{E_b}\right)^{-\alpha+1}, & E < E_b \\ \left(\frac{E}{E_b}\right)^{-\beta+1}, & E > E_{\rm b} \end{cases},
\end{equation}
where $E_b$ is the break frequency measured in the observer frame, $F^{\rm fl}_{E_b} (t)$ is the peak flux at $E_b, $ $\alpha$ and $\beta$ are the spectral indices.
The energy spectrum of EIC emission in the Thomson limit is 
\begin{equation}\label{spec_EIC}
F_E^{\rm EIC} = F_{E, \rm max}^{\rm EIC}  
\begin{cases} \left(\frac{E}{E_m^{\rm EIC}}\right)^{1-\alpha}, & E < E_m^{\rm EIC} \\ \left(\frac{E}{E_m^{\rm EIC}}\right)^{-\frac{(s-1)}{2}}, & E_m^{\rm EIC} < E < E_c^{\rm EIC} \\ \left(\frac{E_c^{\rm EIC}}{E_m^{\rm EIC}} \right)^{\frac{1-s}{2}} \left(\frac{E}{E_c^{\rm EIC}}\right)^{-\frac{s}{2}}, & E > E_c^{\rm EIC} \end{cases},
\end{equation}
where $F_{E, \rm max}^{\rm EIC} \sim \tau_T x F_{E_b}^{\rm fl}$, $F_{E_b}^{\rm fl}$ is the peak flux of the late-prompt emission, and $x < 1$ is a factor due to the anisotropic scattering process~\citep{Murase:2010fq}.

The SSC (EIC) Compton parameter is expressed as the ratio of SSC (EIC) emission power to synchrotron emission power,
\begin{equation}\label{Y}
Y_{\rm SSC(EIC)}(\gamma_e) \approx \frac{P_{\rm SSC(\rm EIC)}}{P_{\rm syn}} \sim \frac{U_{\rm syn(FL)}^\prime[\varepsilon^\prime < \varepsilon_{\rm KN}^\prime]}{U_{\rm B}^\prime},
\end{equation}
where $P_{\rm SSC}$ is the SSC emission power, $P_{\rm EIC}$ is the EIC emission power, $P_{\rm syn}$ is the synchrotron emission power, $U_{\rm syn}^\prime$ is the comoving synchrotron photon energy density, where we introduce the Klein-Nishina energy $\varepsilon_{\rm KN}^\prime \sim m_e c^2 /\gamma_e$ to take into account Klein-Nishina effect, $U_{\rm FL}^\prime$ is the comoving photon density of flares, and $U_B^\prime = B^2 / 8\pi$ is the comoving magnetic energy density.
If we neglect EIC cooling, the value of $Y_{\rm SSC}$ can be estimated to be $Y_{\rm SSC} \sim (\epsilon_e / \epsilon_B)^{1/(4-s)} (\gamma_{e,m} / \gamma_{c, \rm syn})^{-(s-2)/(2(s-4))} \sim 90$ at $t \sim 10^4\rm~s$ in the Thomson regime~\citep{Sari:2000zp, Liu:2013foa}.

It has been shown that both the synchrotron spectrum and the SSC (EIC) spectrum can be affected in the Klein-Nishina regime~\cite[see, e.g.,][]{Nakar:2009er,Wang:2009rp,Murase:2010fq}.
The Klein-Nishina effect on the high-energy IC emission leads to the spectral suppression when the upscattered photons have energies beyond the critical energy~\citep{Blumenthal:1970gc}.
In the observer frame, the characteristic break energies are $E_{\rm KN}^m \approx \Gamma \gamma_{e,m} m_e c^2 / (1+z) \simeq 2.9 \times 10^9~\mathcal{E}_{k,52}^{1/4} n_{\rm ex}^{-1/4}  \epsilon_{e,-1} f_{e}^{-1} t_{z,4}^{-3/4} (1+z)^{-1} \rm~eV$ and $E_{\rm KN}^c \approx \Gamma \gamma_{e,c}  m_e c^2 / (1+z) \simeq 1.5 \times 10^{14}~(1+Y_{\rm tot})^{-1} \mathcal{E}_{k,52}^{-1/4} n_{\rm ex}^{-3/4} \epsilon_{B,-5}^{-1} t_{z,4}^{-1/4} (1+z)^{-1} \rm~eV$.
The Klein-Nishina break energy can be lower than either $E_c^{\rm SSC}$ or $E_c^{\rm EIC}$. The energy spectrum in Equation~\ref{spec_SSC} and Equation~\ref{spec_EIC} can be further affected by the Klein-Nishina effect, where the spectrum is steepened beyond the Klein-Nishina break energy~\citep[see Eqs.~24 and 25 of][]{Murase:2010fq}.

This work focuses on the case that EIC emission dominates over SSC emission because the peak flux of the flare emission is much larger than the afterglow synchrotron emission, i.e., $F_{E_b}^{\rm FL} > F_{E, \rm max}^{\rm syn}$.
However, the evolution of the blast wave may still be in the coasting phase during the stage of the flare emission.
In our numerical calculations, we consider the coasting phase, deceleration phase and non-relativistic phase by solving a series of partial differential equations. See Appendix~\ref{dynamics} for details.
Note the emission during the non-relativistic evolution phase is necessary for explaining the late-time radio data~\citep{Rhodes:2020hsx}.
By extending the method adopted in~\cite{Murase:2010fq}, we numerically calculate spectra of synchrotron, SSC, and EIC emission. We take into account the equal-arrival-time surface (EATS) for not only the EIC component but also the synchrotron and SSC components. See details in Appendix~\ref{radiative}.

\section{Modeling high-energy gamma-ray emission from \grb}\label{sec:grb}
Following~\cite{Chand:2020wqt}, we fit the observed X-ray light curve with two separate components, the late-time flare emission and external forward shock emission, respectively.
The X-ray light curve drops very quickly during the flare stage, which is difficult to explain in the standard afterglow model with typical electron spectral indices of $s \sim 2 - 2.5$.
The late-time X-ray flare can be fitted with the Norris model~\citep{Norris:2005ew, Chand:2020wqt},
\begin{equation}
F^{\rm fl}_{\varepsilon_b}(t) = A \lambda e^{-\frac{\tau_1}{t-t_i} - \frac{t-t_i} {\tau_2}},
\end{equation}
where $A$ is the pulse amplitude, $t_i$ is the pulse start time, $\tau_1$ is the pulse r, andise parameter $\tau_2$ is the pulse decay parameter, $\lambda$ is defined as the normalization constant $\lambda = {\rm exp}[2(\tau_1/\tau_2)^{1/2}]$. 
The best-fit results of the above parameters are $A = 225 \rm~\mu Jy$, $\tau_1 = 90\rm~s$, $\tau_2 = 3993\rm~s$, and $t_i = 950\rm~s$. 
The corresponding values used in Equation~\ref{eq:fl} are $\alpha = 1$, $\beta = 2.5$ and $E_b = 100\rm~eV$, which is optimized for brighter EIC fluxes.

\begin{figure}
\includegraphics[width=\linewidth]{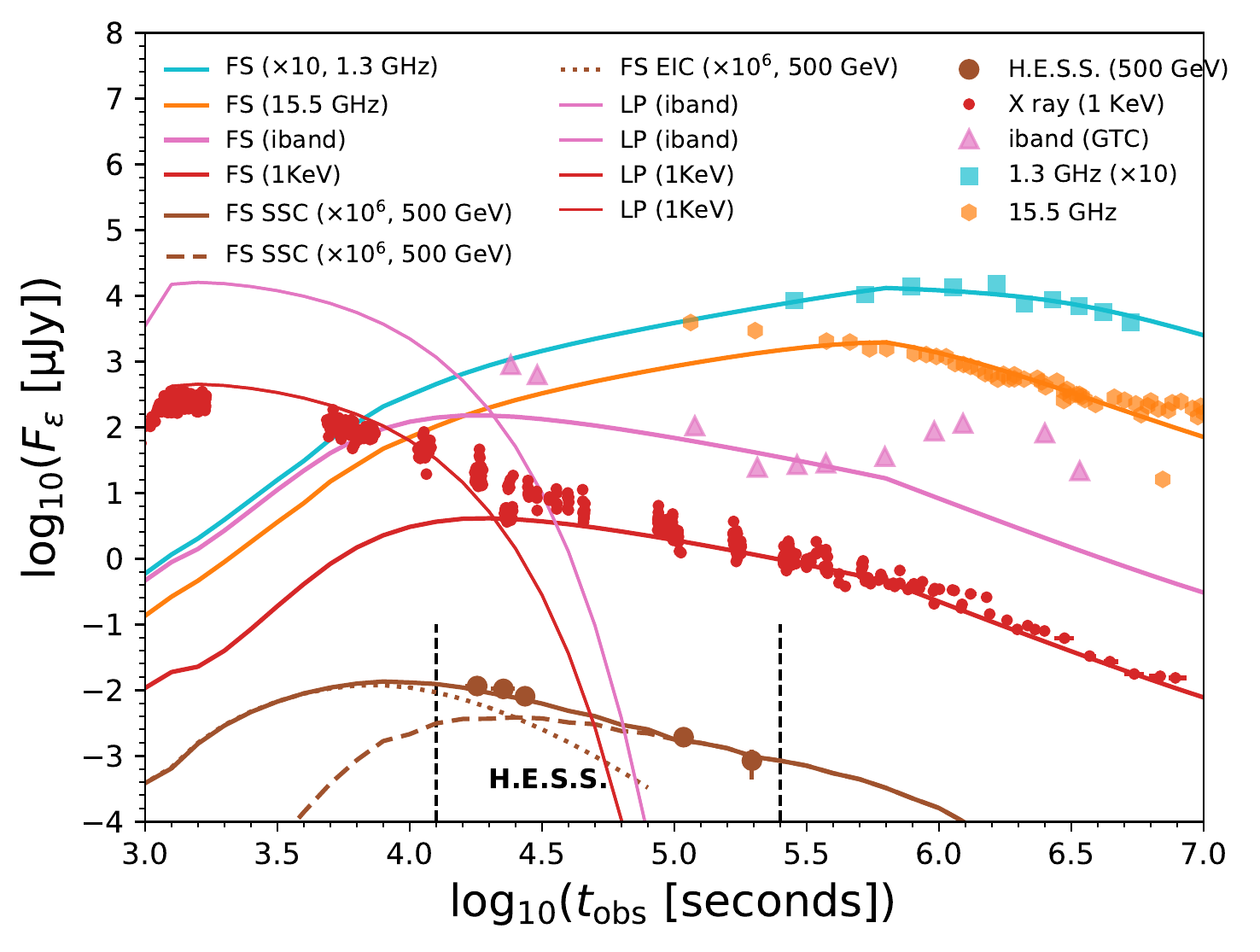}
\caption{Multi-wavelength light curve from radio, optical, X-ray to TeV bands. We show both the forward shock (FS) light curve (thick lines) and the late-prompt (LP) light curve (thin lines). The H.E.S.S. observation time window is indicated within black dashed vertical lines. The relevant physical parameters are $\mathcal{E}_k = 9.8 \times 10^{51}\rm~erg$, 
%$\mathcal{E}_k = 9.78 \times 10^{51}\rm~erg$
$n_{\rm ex} = 0.09\rm~cm^{-3}$, $\epsilon_e = 0.39$, $f_e = 0.34$, $\epsilon_B = 8.7 \times 10^{-5}$, $s=2.1$, $\theta_j = 0.2$, $\Gamma_0 = 25$, $\alpha = 1$, $\beta = 2.5$ and $E_b = 100\rm~eV$}. 
\label{fig:GRB190829A_Model4_light curve}
\end{figure}

In Fig.~\ref{fig:GRB190829A_Model4_light curve}, we show the multi-wavelength light curve from radio, optical, X-ray to VHE band.
The X-ray light curve at 1 keV that has been observed by \textit{Swift-XRT} is taken from the public online repository\footnote{We convert the observed flux at 0.3-10 keV to that at 1 keV, assuming a spectral index of $\Gamma_X = 1.8$.}~\citep{Evans2010}. 
We can see the observed X-ray flux is well explained with both contributions from the late-prompt flare and forward shock emission.
Assuming the opening angle of the outflow is $\theta_j \sim 0.2$ radian, the jet break occurred at $t \sim 10^5\rm~s$. Although there is a lateral expansion, the spectral decline after the jet break is known to be dominated by the geometrical effect, so we simply multiply a correction factor $\theta_j^2 \Gamma^2$ to the observed flux reduction after the jet break~\citep{Zhang:2018ond}.
The pink triangles are the optical $i$-band data observed by the Gran Telescopio CANARIAS (GTC) after the correction for both Galactic and host galaxy extinction~\citep{Hu:2020xhu}.
As done for X-rays, we fit the optical light curve as a combination of the late-prompt flare and external forward shock emission. Note that the bump that appeared in the late-time optical data should be attributed to supernova emission.
We also show the radio data at 1.3 GHz band (yellow squares) and 15.5 GHz band (blue points) observed by Meer Karoo Array Telescope (MeerKAT) and Arcminute Microkelvin Imager - Large Array (AMI-LA) one day after the burst, respectively~\citep{Rhodes:2020hsx}.
In our model, the radio light curve at 1.3 GHz and 15.5 GHz can be explained in the external forward shock model, even though possible contributions may also come from the external reverse shock~\citep{Rhodes:2020hsx}.
Note that we include an additional parameter $f_e$ in our model to consider the case where only a fraction of the electrons are injected into the acceleration process, which can affect the minimum electron Lorentz factor $\gamma_{e,m}$ as well as the flux at the radio band~\citep{Samuelsson:2020upt}.

We show the SSC and EIC light curves at 500 GeV as dashed and dotted lines, respectively.
The EIC light curve dominates the high-energy emission at $t \sim 10^3 - 10^4\rm~s$, and gradually declines due to the decrease of the late-prompt flare emission. The SSC light curve becomes dominant at later times, $t\gtrsim 3 \times 10^4\rm~s$.
We define $\theta_{\rm sc}^\prime$ as the scattering angle relative to the photon beam measured in the external forward shock comoving frame. 
The EIC emission diminishes at $\theta_{\rm sc}^\prime = 0$. See Appendix~\ref{radiative} for more details. 
The EIC emission nearly coincides with the late-prompt flare, but it extends a factor of $\sim 3-4$ times longer than the duration of the flare~\citep{Murase:2010fq}.
Despite the rapid evolution of the late-prompt flare, the EIC flux decreases slowly due to the effect of the integration over the EATS. Note that for a given time the angle at which non-thermal electrons produce the dominant fraction of the EIC flux is $\theta \lesssim \Gamma^{-1}$~\citep[]{Murase:2010fq}, which is smaller than the jet opening angle $\theta_j$ before the jet break.

\begin{figure}
\includegraphics[width=\linewidth]{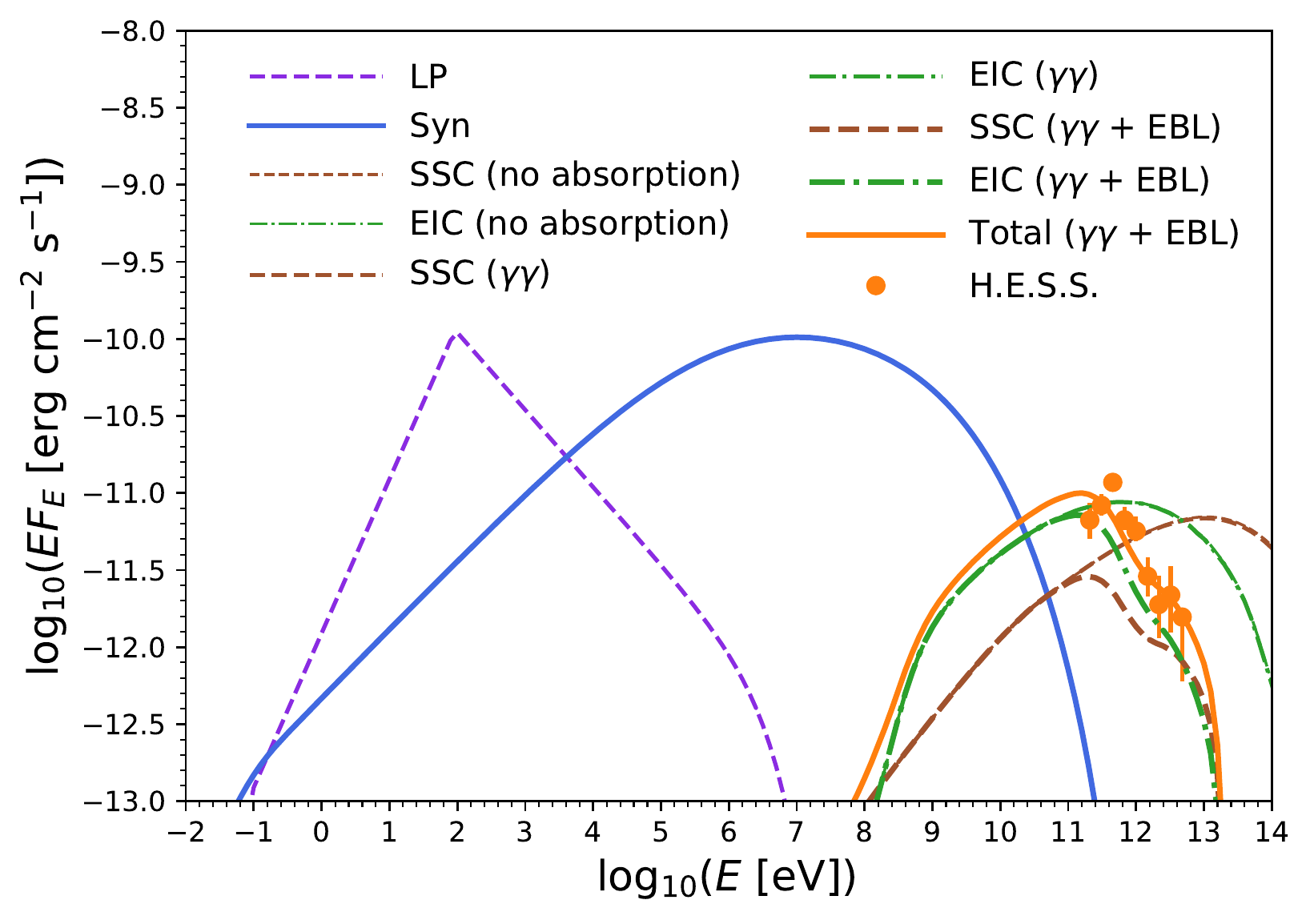}
\caption{Same as Fig.~\ref{fig:GRB190829A_Model4_light curve}, we show energy spectra of the synchrotron, SSC, EIC, and late-prompt (LP) flare emission at $t = 4.4 \rm~hr$. The thick green, red and yellow lines take into account both of the internal $\gamma \gamma$ absorption and EBL attenuation. The H.E.S.S. VHE spectra data on the first night are taken from ~\citet{HESS2021}.
%while the thin lines ignore the EBL attenuation.
\label{fig:spectrum}}
\end{figure}

In Fig.~\ref{fig:spectrum}, we show energy spectra of various components including the flare, synchrotron, SSC, and EIC at t = $3.5\rm~hrs$.
The synchrotron emission peaks around $E_c \sim 10^6 \rm~eV$ and the characteristic energy is $E_m \sim 10^{-1}\rm~eV$. 
It is clear that the upscattering of these higher-energy photons is limited by the Klein-Nishina effect. The maximum EIC flux is observed at $t \sim 3.5\rm~hr$ with a peak flux of $\sim 10^{-10.5}\rm~erg~cm^{-2}~s^{-1}$. At the same time, the SSC flux is $\sim 5$ times lower than the EIC flux, as shown in Fig.~\ref{fig:spectrum}.
The cutoff of the VHE emission at $\gtrsim 500\rm~GeV$ is dominated by the EBL attenuation during their propagation from the source to Earth. Note that the effect of the Klein-Nishina break energy is difficult to observe due to the strong EBL attenuation.

\vhe from \grb have been detected by \hess with a confidence level $>5\sigma$~\citep{2019ATel13052....1D}.
The observations started at $\sim 1.5\times 10^4\rm~s$ after the GRB trigger, where the integral flux beyond 200 GeV reached $\sim4 \times 10^{11}\rm~erg~cm^{-2}~s^{-1}$ and gradually decline following the same trend as X-rays until $t \sim 2 \times 10^5\rm~s$~\citep{HESS2021}.
Our calculations predict that the high-energy EIC emission has a flux of $\gtrsim 10^{-11} \rm~erg~cm^{-2}~s^{-1}$ consistent with the \hess observation in the earlier observation time, while the late time observation at $t \sim 10^5\rm~s$ can be explained by the SSC emission.

\section{Discussion}
\begin{figure}
\centering
\includegraphics[width=\linewidth]{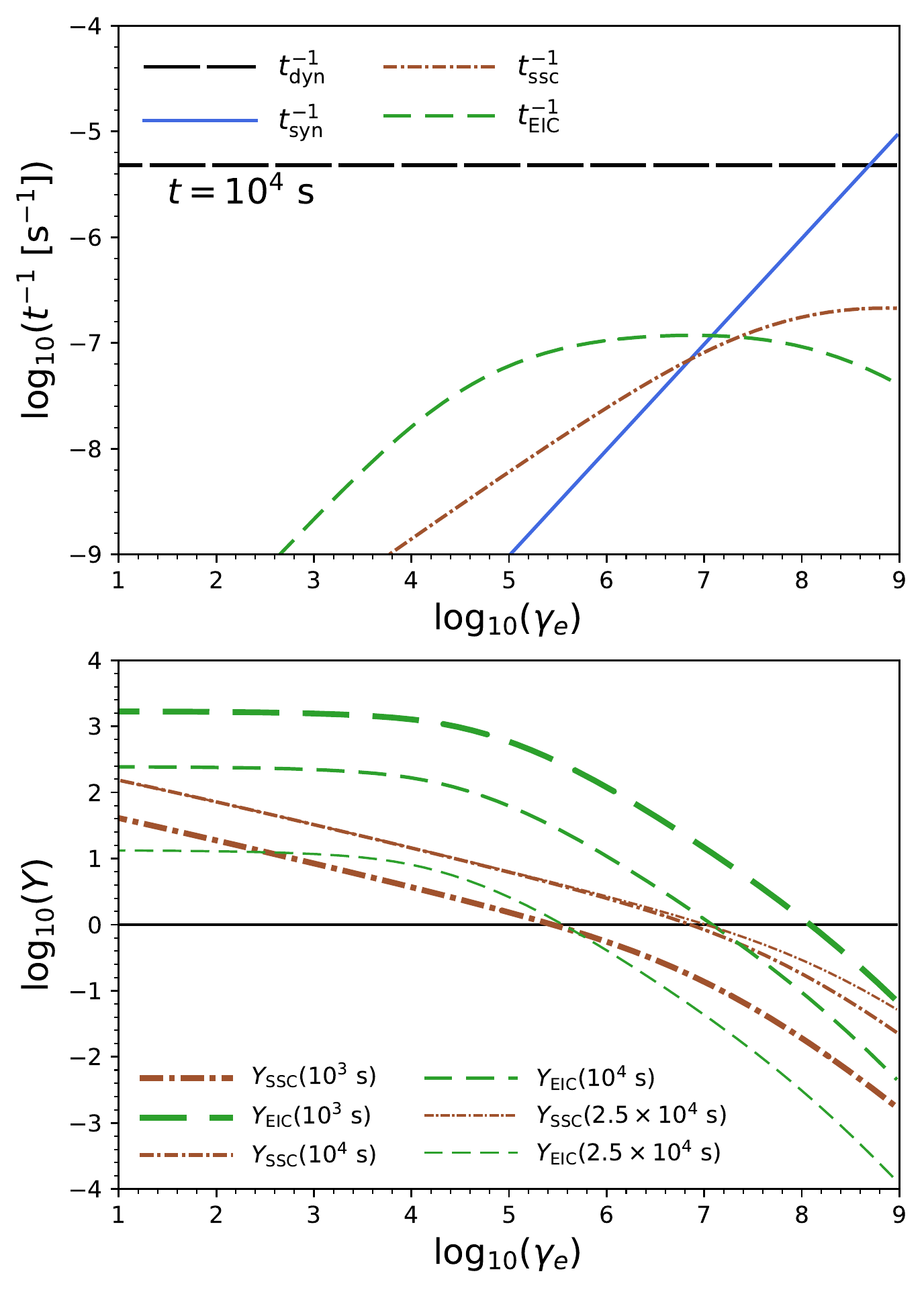}
\caption{\textbf{Upper panel}: Dynamical timescale and various electron cooling timescales for synchrotron, SSC and EIC as a function of electron Lorentz factor $\gamma_e$. \textbf{Lower panel}: The evolution of $Y_{\rm SSC}(\gamma_e)$ and $Y_{\rm EIC}(\gamma_e)$ as a function of $\gamma_e$ at $t = 10^3\rm~s$, $t = 10^4\rm~s$ and $t = 2.5 \times 10^4\rm~s$.}
\label{fig:timescale}
\end{figure}

\vhe from \grb have been studied based on the SSC scenario, as in other VHE GRBs~\citep{Chand:2020wqt, Fraija:2020vsa, Rhodes:2020hsx, Hu:2020xhu, Sato2021}. However, the relative ratio of the EIC component to SSC component depends on both the energy spectrum of flare emission and synchrotron emission, $F_{E = 500 \rm GeV}^{\rm EIC} / F_{E = 500 \rm GeV}^{\rm SSC} = (F_{E_b}^{\rm fl} / F_{E, \rm max}^{\rm syn}) (E_m / E_b)^{-(s-1)/2} \sim Y_{\rm EIC} / Y_{\rm SSC}$ which can be derived from Equations~\ref{spec_SSC}, \ref{spec_EIC} and \ref{Y}.
In Fig.~\ref{fig:timescale}, we show the synchrotron cooling timescale $t_{\rm syn}$, the SSC cooling timescale $t_{\rm SSC}$, and the EIC cooling timescale $t_{\rm EIC}$, as a function of $\gamma_e$ at $t = 10^4\rm~s$.
We also show the time evolution of the Compton parameters, $Y_{\rm SSC}(\gamma_e)$ and $Y_{\rm EIC}(\gamma_e)$, at various observation times. 
We can see that $Y_{\rm SSC}(\gamma_e)$ is not constant and it declines with the increase of $\gamma_e$ due to the Klein-Nishina effect. 
According to Equation~\ref{Y}, the value of $Y_{\rm SSC}(\gamma_e)$ is proportional to the comoving synchrotron photon energy density $U_{\rm syn}^\prime [\varepsilon^\prime < \varepsilon_{\rm KN}^\prime] \propto {\varepsilon_{\rm KN}^\prime}^{1/2}$.
Note $\varepsilon_{\rm KN}^\prime$ is proportional to $\gamma_e^{-1}$, we can expect $Y_{\rm SSC}(\gamma_e) \propto \gamma_e^{-1/2}$ consistent with $Y_{\rm SSC}(\gamma_e)$ shown in Fig.~\ref{fig:timescale}.
Unlike $Y_{\rm SSC}(\gamma_e)$, the Compton parameter $Y_{\rm EIC}(\gamma_e)$ remains constant up to $\gamma_e \sim 10^5$. 
The reason is that the energy density of the late-prompt photons $U_{\rm EIC}^\prime [\varepsilon^\prime < \varepsilon_{\rm KN}^\prime]$ remains constant, which is dominated by the energy density near $\varepsilon_b \sim 100\rm~eV$ as long as $\gamma_e \lesssim 10^4$. 
One visible feature of $Y_{\rm EIC}(\gamma_e)$ is the rapid decline following the time evolution of the X-ray flare.

The physical parameters of LL GRBs are still uncertain due to the limited number of observations and the values used in Fig.~\ref{fig:GRB190829A_Model4_light curve} are optimized for fitting the multi-wavelength light curve.
The values of the microphysical parameters are $\epsilon_e \sim 0.4$, $f_e \sim 0.3$, and $\epsilon_B = 9 \times 10^{-5}$, assuming uniform microturbulence in the shock vicinity.
For GRB afterglows, the typical value of $\epsilon_e$ is $\sim 0.1$, while the value of $\epsilon_B$ varies in a wide range, e.g., $\sim 10^{-5} - 10^{-1}$~\citep{Kumar:2014upa}.
Small values of $\epsilon_B$ can be consistent with ones expected around the contact discontinuity, considering the decay of $\epsilon_B$ from a stronger value of $\sim 0.01$ at the shock front inferred from particle-in-cell simulations ~\citep{Lemoine:2012yw, Vanthieghem:2020nvr}.

\begin{figure}
\includegraphics[width=\linewidth]{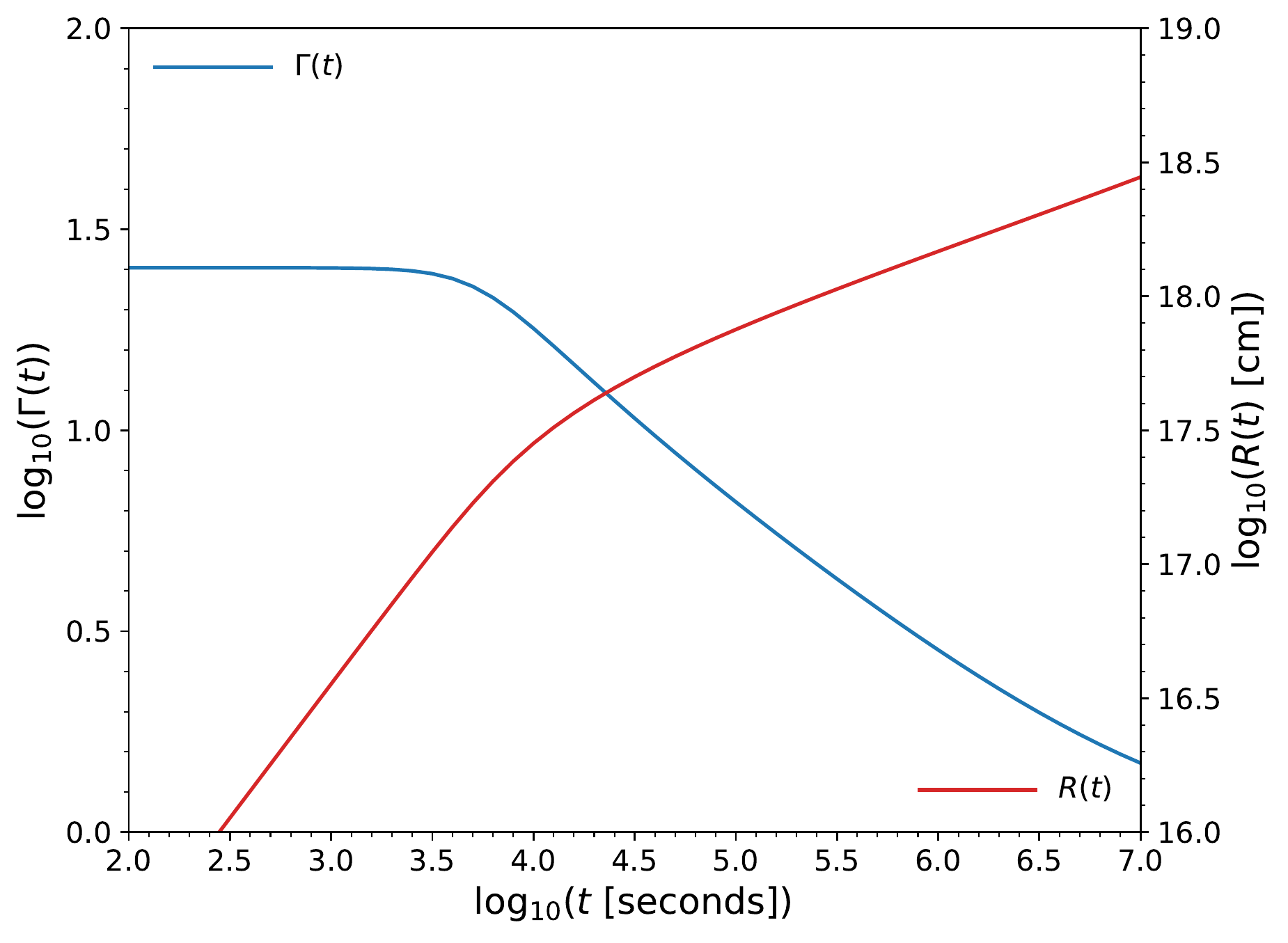}
\caption{Time evolution of the Lorentz factor and radius of the blast wave as a function of the observation time since the burst, $t$.
\label{fig:dynamic}}
\end{figure}

In this work, we consider an impulsive relativistic outflow with a kinetic energy of $\mathcal{E}_k = 1 \times 10^{52}\rm~erg$ and an initial Lorentz factor of $\Gamma_0 = 25$, propagating into an external medium with a constant density of $n_{\rm ex} = 0.14\rm~cm^{-3}$.
The value of kinetic energy and $\Gamma_0$ is larger than for LL GRBs, e.g. GRB 980425~\citep{Galama:1998ea, Kulkarni:1998qk} and GRB 060218~\citep{Soderberg:2006vh, Campana:2006qe}, but
smaller than canonical high-luminosity GRBs~\citep{Kumar:2014upa}.
The value of $\Gamma_0$ should not be much smaller, e.g. $\Gamma < 10$, otherwise the deceleration time is too long. 
A higher value of $\Gamma_0 > 25$ will enhance the SSC contribution at earlier time $t\lesssim 10^4\rm~s$, but the flux of \vhe still dominated by EIC components. 
We stress that the advantage of the EIC+SSC model presented in this work is that in the presence of X-ray flares it gives a better fit to the multi-wavelength light curve with microphysical parameters that are similar with other TeV GRBs.
In Fig.~\ref{fig:dynamic}, we show the time evolution of the Lorentz factor and radius of the blast wave.
The deceleration time can be determined when the total mass of the swept-up matter equals to a fraction $1/\Gamma_0$ of the ejecta mass, $R_{\rm dec} \simeq 2.9 \times 10^{17}\mathcal{E}_{k, 52}^{1/3} \Gamma_{0,1.4}^{-2/3} n_{\rm ex, -1}^{-1/3}\rm~cm$ and the corresponding deceleration time can be estimated to be $t_{\rm dec} \approx (1 + z) R_{\rm dec}/2 \Gamma_0 c^2 \simeq  8.4 \times 10^{3} (1 + z) \mathcal{E}_{k, 52}^{1/3} \Gamma_{0,1.4}^{-8/3} n_{\rm ex,-1}^{-1/3}\rm~s$.
In addition, we note that the predicted Lorentz factor of the ejecta is $\Gamma \sim 1.7$ at $\sim 60$ days which is only slightly smaller than the velocity measured by the Very Long Baseline Interferometry (VLBI)~\citep{GirolettiVLBI}.

Our results can also be affected by the energy spectrum of late-prompt emission. For example, the peak flux $F_{\varepsilon_b}$ used in Equation~\ref{eq:fl} will decrease for a smaller value of $\beta$, given the constraints at the X-ray band.
In this work, the break energy $\varepsilon_b$ of the late-prompt flare emission is chosen to be constant. In general, the value of $\varepsilon_b$ may decrease with time~\citep{Murase:2010fq}. However, the evolution of $\varepsilon_b$ is not essential for our conclusions, since lower values of $\varepsilon_b$ at later times can enhance the scattering efficiency in the Thomson regime when $\gamma_{e,c}$ becomes larger.
Compared to canonical high-luminosity GRBs, energy spectra of LL GRBs usually have lower $\varepsilon_{m}$ and higher $\varepsilon_c$, with the dominance of photons with energies around $\varepsilon_c$. 
The Klein-Nishina effect in the SSC spectrum will be more relevant in such a situation.
The IC scattering between the same population of electrons with external photons can alleviate the Klein-Nishina suppression if the external photons have the proper energies that can keep the IC scattering process in the Thomson regime.

\grb has the lowest radio luminosity in the GRB samples considered in~\cite{Rhodes:2020hsx}. However, there is no difference in the radio luminosity among all of the three VHE GRBs and other GRBs. 
The associated supernova SN 2019oyw appeared to have similar ejecta mass and kinetic energy as SN 1998bw~\citep{Hu:2020xhu} that was associated with a LL GRB. 
Note that thermal photons from an associated supernova can also be upscattered by non-thermal high-energy electrons accelerated by the conventional GRB outflow or trans-relativistic ejecta~\citep{Asano:2008rk, Ando:2008yf, He:2009rx}.
Hadronic \vhe from LL GRBs have also been studied. In particular, \cite{Murase:2008mr} and \cite{Murase:2010va} proposed heavy-ion synchrotron emission and \vhe from the deexcitaion of the UHECR nuclei, respectively~\citep[see, e.g., Fig.~9 of][]{Murase:2008mr}. On the other hand, \cite{Murase:2011yw} and \cite{Kashiyama:2012zn} suggested hadronic \vhe originating from $p\gamma$ interactions following prompt and shock breakout emission, respectively. 
These mechanisms would compete with high-energy emission from primary electrons~\citep{Ando:2008yf}. 
The detailed modeling of the hadronic processes and the following electromagnetic cascade is essential~\citep[see, e.g.,][as an example of hadronic and leptonic components]{Murase:2010va} to disentangle the hadronic component from the leptonic component and to obtain constraints both on cosmic-ray acceleration mechanisms and on radiation processes.

\section{Summary}\label{sec:summary}
In this work, we showed that \vhe are naturally produced by the EIC+SSC mechanism, which may provide an explanation for the observation of \vhe from \grb by \hess. The EIC scenario is seeded by the observed long-lasting late-prompt flare in the X-ray and optical bands as target photons. Assuming that the non-thermal electrons are accelerated in the external forward shock, these can upscatter the late-prompt flare photons, and our results show that the EIC components can contribute to the \hess observations at $t \sim 1 \times 10^4\rm~s$. 

We consider an impulsive relativistic outflow propagating into an external medium with a constant density, where the physical parameters adopted in this work are optimized for fitting the multi-wavelength light curve.
We showed that our model can explain the multi-wavelength observations of \grb, given $\mathcal{E}_k = 10^{52}\rm~erg$, $\Gamma_0 = 25$, and $n_{\rm ex} \sim 0.1\rm~cm^{-3}$. Our results suggest that \grb is not a typical LL GRB but has much higher kinetic energy.
The TeV photons observed by \hess can be dominated by EIC emission at earlier times, while the SSC component becomes dominant at later times.
Our results suggest that the EIC emission nearly coincides with the late-prompt flare, but decreases more slowly than the evolution of the flare emission, and extends $\sim 3-4$ times longer than the duration of the flare emission.
Future observations of EIC components that are related to late-prompt flares in GRBs will be helpful for constraining the properties of the outflow and central engine activities.

\begin{acknowledgements}
The work of K.M. is supported by the Alfred P. Sloan Foundation, NSF Grant No.~AST-1908689, and KAKENHI No.~20H01901 and No.~20H05852. 
B.T.Z. acknowledges the IGC fellowship. 
P.M. acknowledges support from the Eberly Foundation. 
P.V. acknowledges support from NASA grants 80NSSC19K0595 and NNM11AA01A. 
\end{acknowledgements}

\bibliographystyle{aasjournal}
\bibliography{bzhang}

%\clearpage

\appendix
We calculate time-dependent spectra of high-energy afterglow emission by extending the method used in \cite{Murase:2010fq}. We numerically calculate afterglow dynamics to treat not only the relativistic regime (described by the self-similar solution) but also the non-relativistic regime, and take into account effects of EATS.

\section{Afterglow dynamics}\label{dynamics}
We consider an impulsive relativistic outflow with kinetic energy $\mathcal{E}_k$ and initial Lorentz factor $\Gamma_0$, propagating into an external medium of constant density $n_{\rm ex}$.
The total energy of the blast wave which develops is given by~\citep{Nava:2012hq}

\begin{equation}
\mathcal{E}_{\rm tot} = \Gamma M_{\rm ej} c^2 + \Gamma m c^2 + \frac{\hat{\gamma}\Gamma^2 - \hat{\gamma} + 1}{\Gamma} \mathcal{E}_{\rm int}^\prime,
\end{equation}

where $M_{\rm ej} = \mathcal{E}_k/\Gamma_0 c^2$ is the mass of the outflow, $m = (4\pi/3) r^3 n_{\rm ex} m_p$ is the mass of the swept-up external matter, $\hat{\gamma} = (4+\Gamma^{-1})/3$ is the adiabatic index which is a good approximation in both relativistic and non-relativistic regimes~\citep{Nava:2012hq}. 
The comoving internal energy is $\mathcal{E}_{\rm int}^\prime=(\Gamma - 1) m c^2$.
Considering $d\mathcal{E}_{\rm tot} = dm c^2$ and neglecting both adiabatic and radiative energy losses \citep[see][for details]{Nava:2012hq, Zhang:2018ond}, we can derive the following 1D differential equation,

\begin{equation}\label{Gamma}
\frac{d\Gamma}{dm} = -\frac{\Gamma - 5\Gamma^3 + 4\Gamma^5}{3 M_{\rm ej} \Gamma^3 - 2 m + 8\Gamma^4 m}.
\end{equation}

The differential mass of the collected external medium is $dm = 4\pi r^2 n_{\rm ex} f_{\rm corr} m_p dr$. We can derive the well-known relation $\Gamma \propto r^{-2/3}$ once the blast wave enters into the Blandford-McKee (BM) self-similar adiabatic evolution regime~\citep{Blandford:1976uq}. 
Note that we multiply a factor of $f_{\rm corr} = 9/17$ to the swept-up external matter density in order to match the normalization of the BM self-similar solution~\citep{Nava:2012hq}. 
The BM self-similar phase begins after a significant deceleration occurs, where the kinetic energy of the initial ejecta equals the sum of the kinetic energy of the swept-up external matter and its internal energy $\Gamma M_{\rm ej} c^2 \sim \hat{\gamma}\Gamma ^2 m c^2$. 
In order to obtain the value of quantities in the observer frame, we adopt the differential relation, $dr=\beta c dt / (1 - \beta)$, which represents the case that photons propagate with distance $dr$ can be observed in a time interval $dt$.
The differential equation derived in Equation~\ref{Gamma} also gives an appropriate description of the blast wave evolution in both the coasting ($\Gamma = \Gamma_0, r \propto t$) and non-relativistic deceleration ($\beta \propto t^{-3/5}, r \propto t^{2/5}$) regimes~\citep{Huang:1999di, Panaitescu:2000bk, Peer2012, Nava:2012hq, Lu:2020hka}.

\section{Non-thermal electron distribution}\label{electrons}
\begin{figure}
%\centering
\includegraphics[width=0.49\textwidth]{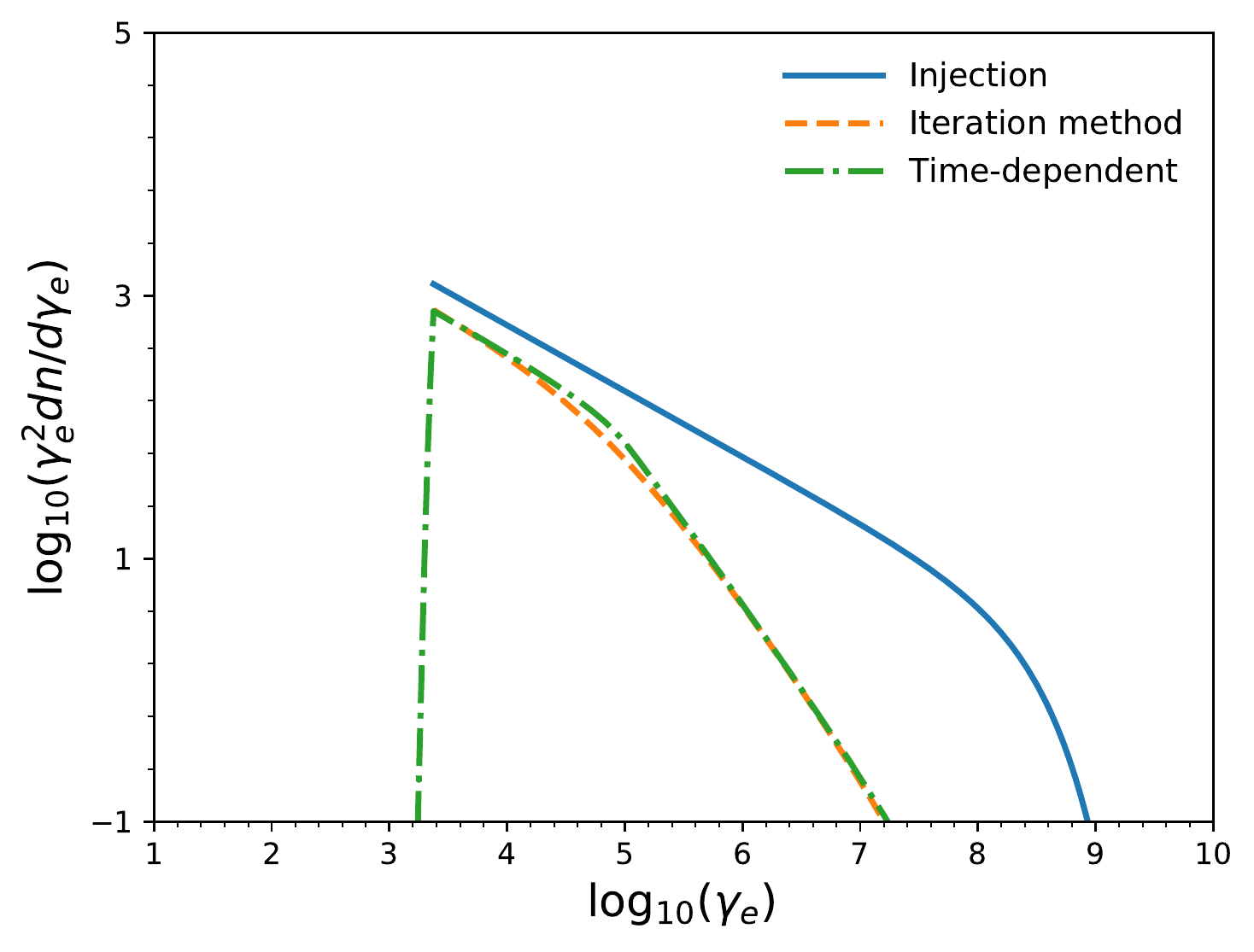}
\includegraphics[width=0.49\textwidth]{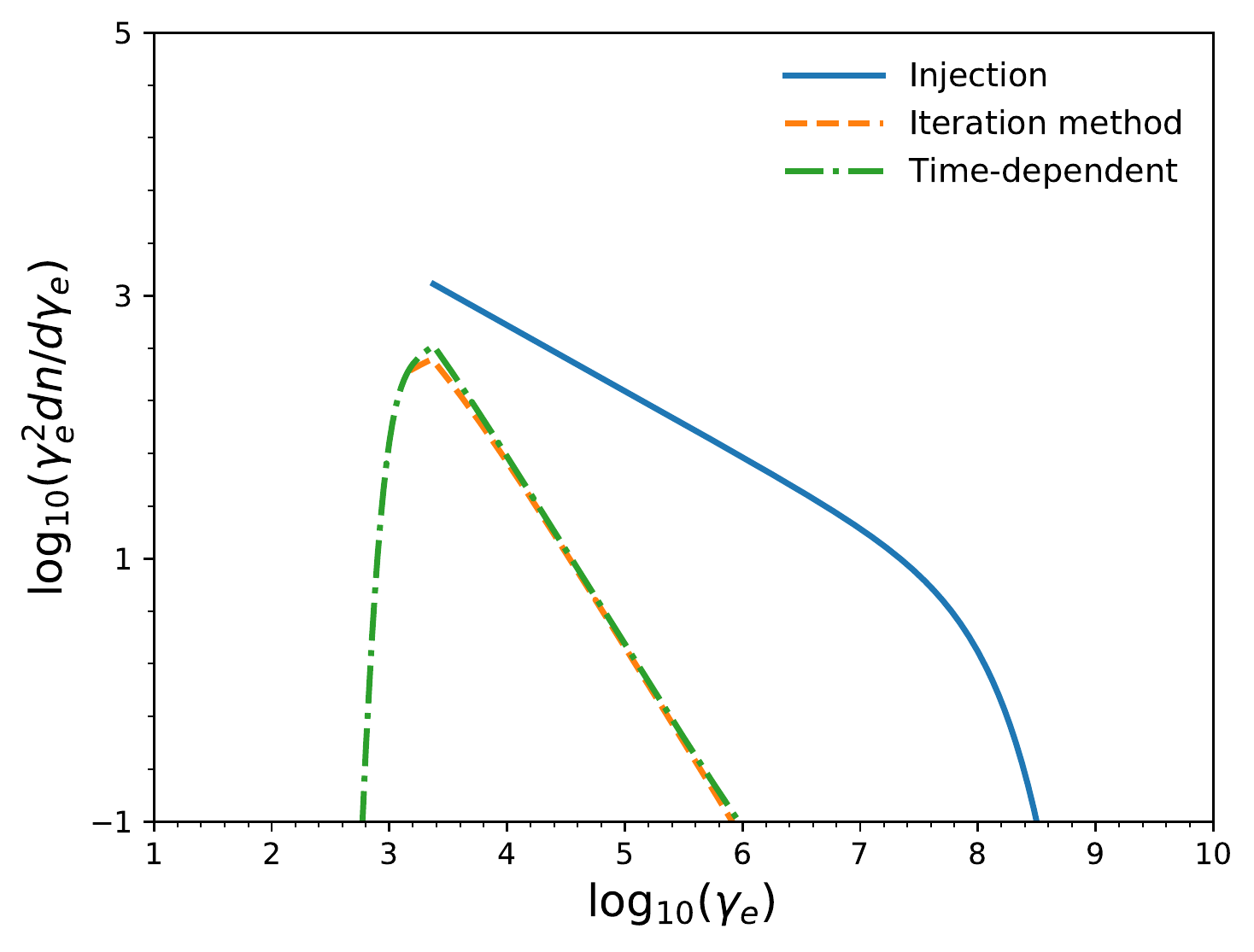}
\caption{ Comparison of the electron energy spectra derived by the  iteration method used for the main results and by solving the kinetic equation at $t' = 100\rm~s$. We use $\mathcal{E}_k = 1 \times 10^{52}\rm~erg$, $n_{\rm ex} = 1\rm~cm^{-3}$, $\epsilon_e = 0.3$, $f_e = 1.$, $s=2.5$, $\Gamma_0 = 50$. 
\textit{Left panel}: slow-cooling regime, $\epsilon_B = 10^{-3}$. \textit{Right panel}: fast-cooling regime, $\epsilon_B = 10^{-1}$.}
\label{electron_compare}
\end{figure}

We calculate afterglow synchrotron and SSC spectra numerically, and we confirm that the numerical results agree with the analytical results. 
In general, electron energy spectra in the downstream of the external forward shock can be derived by solving the kinetic equation~\citep{Blumenthal:1970gc},
\begin{equation}\label{continuity-equation}
\frac{\partial n_{\gamma_e}(t')}{\partial t'} + \frac{\partial}{\partial \gamma_e} \left(n_{\gamma_e}(t') \dot{\gamma}_e \right) + \frac{n_{\gamma_e}(t')}{t'_{\rm esc}}=\dot{n}_{\gamma_e}^{\rm inj}(t'),
\end{equation}
where $n_{\gamma_e}(t^\prime)$ is the number density of electrons per electron Lorentz factor, $\dot{\gamma}_e = d\gamma_e/dt^\prime = \gamma_e {t^\prime}_{\rm cool}^{-1}$ is the electron energy loss rate (divided by $m_ec^2$), $t^\prime_{\rm esc}$ is the possible escape time, $\dot{n}_{\gamma_e}^{\rm inj}(t')=\mathcal{C} \gamma_e^{-s}$ is the electron injection rate where $\mathcal{C}\approx (s-1) \gamma_{e,m}^{s-1} n_e /t'_{\rm dyn}$ (for $s>2$), $\gamma_{e,m}$ is the electron minimum Lorentz factor, $n_e$ is the non-thermal electron number density in the comoving frame  which is normalized based on $N_e = 4\pi r^2 n_e t'_{\rm dyn} c = (4\pi/3) r^3 n_{\rm ex} f_e$, and considering an on-axis observer we use $t^\prime_{\rm dyn}\approx\Gamma t/(1+z)$ as the dynamical timescale. 
The electron minimum Lorentz factor is given by $\gamma_{e,m} \approx (\epsilon_e /f_e)[(s-2)/(s-1)](m_p/m_e)(\Gamma-1)$ for $s>2$, where $\epsilon_e$ is the energy fraction of thermal energy that goes into non-thermal electrons and $f_e$ is the number fraction of thermal electrons that are injected into the acceleration process.
The total cooling rate consists of radiative cooling and possible adiabatic losses. 
The synchrotron loss rate is ${t^\prime}_{\rm syn}^{-1}= (4/3)\sigma_Tc(\gamma_e/m_e c^2)\beta^2 U_{\rm B}$, where $U_B = B^2/8\pi$ is the magnetic energy density~\citep{Rybicki:1986}. 
The IC loss rate ${t^\prime}_{\rm IC}^{-1}$ is implemented as in Equation~(B3) of \cite{Murase:2010fq}, which includes both SSC and EIC processes. 
The adiabatic loss rate is ${t^\prime}_{\rm ad}^{-1}\sim {t^\prime}_{\rm dyn}^{-1}$. 
In general, the electron distribution is obtained by solving the above differential equation in a time-dependent manner. We call this method ``numerical, time-dependent", but we demonstrate the results using $t'_{\rm esc}=t'_{\rm dyn}$, by which we approximately take into account adiabatic energy losses or particle escape after the decay of magnetic fields in the downstream.
The steady-state solution via $\partial n_{\gamma_e}/\partial t'=0$ can also be found as in Equation~(C.11) in \citet{Dermer:2009zz}. 
On the other hand, for results presented in the main text, we take the iteration method as in \citet{Murase:2010fq}. However, different from the previous work that calculated $\gamma_{e,c}$ with analytical synchrotron spectra and used broken power-law electron distributions, we determine the electron distribution using numerical synchrotron spectra through the following function,
\begin{equation}
n_{\gamma_e}(t') = \frac{1}{{t^\prime}_{\rm dyn}^{-1}+{t^\prime}_{\rm cool}^{-1}}\frac{1}{\gamma_e}\int d\gamma_e^\prime \dot{n}_{\gamma_e^\prime}(t').
\end{equation}
This function is motivated by the steady-state solution for ${t^\prime}_{\rm cool}^{-1}={t^\prime}_{\rm syn}^{-1}+{t^\prime}_{\rm IC}^{-1}+{t^\prime}
_{\rm ad}^{-1}$ in the no escape limit. In the fast cooling case, the electron distribution is mostly described by the steady-state distribution for ${t^\prime}_{\rm cool}^{-1}={t^\prime}_{\rm syn}^{-1}+{t^\prime}_{\rm IC}^{-1}$. 
%as in \citet{Murase:2010va}. 
In the slow cooling case, it is essentially the injection distribution with spectral steepening by radiative cooling. This method was also used in \citet{Asano:2020grw} and \citet{Zhang:2020tem}. 
In Fig.~\ref{electron_compare}, we compared the electron energy spectrum derived using iteration method mentioned above and the spectrum via numerically solving time-dependent kinetic equation Equation~\ref{continuity-equation}.

For comparison, we will also show analytical afterglow synchrotron and SSC spectra, and this method is called ``analytical'', where $\gamma_{e,c}$ is determined by the iteration method to evaluate the Compton $Y$ parameter.

\section{Radiative processes}\label{radiative}
The observed radiation flux from a relativistically moving object can be derived via integration over the equal-arrival-time surface (EATS)~\citep[e.g.,][]{Granot:1998ep,Woods_1999} 
\begin{equation}\label{eq:EATS}
F_E (t) = \frac{1+z}{d_L^2}\int_0^{2\pi} d\phi \int_{-1}^1 d\mu \int_0^\infty dr r^2 \frac{j_{\varepsilon^\prime}(\varepsilon^\prime, \Omega^\prime, \hat{t}, \textbf{r})}{\Gamma^2 (1 - \beta {\rm cos}\theta)^2},
\end{equation}
where $z$ is the source redshift, $d_L$ is the luminosity distance of the source, $j_{\varepsilon^\prime}(\varepsilon^\prime, \Omega^\prime, \hat{t}, \textbf{r})$ is the comoving emissivity at $\textbf{r}$ and time $\hat{t}$, $\mu = {\rm cos}\theta$, $\Gamma$ is the Lorentz factor. The observation time is $t = (1+z)(\hat{t} - r\mu/c)$, where we assume $t = 0$ is the arrival time of photon emitted at origin at $\hat{t} = 0$.
We assume instantaneous emission at $\hat{t}_i$, $j_{\varepsilon^\prime} = j_{\varepsilon^\prime} \delta (\hat{t} - \hat{t}_i) \Delta \hat{t}_i = j_{\varepsilon^\prime} \delta (\hat{t} - \hat{t}_i) (r_i / \beta c\Gamma)$.
Using the relation $\delta (\hat{t} - \hat{t}_i) = \delta (\hat{\mu} - \hat{\mu}_i) c/r_i$, Equation~\ref{eq:EATS} can be simplified as
\begin{equation}
F_E (t) = \frac{(1+z) 2\pi}{d_L^2} \int_0^\infty dr r^2 \frac{j_{\varepsilon^\prime} (\varepsilon^\prime, r, \hat{t})}{\Gamma^3 \beta (1 - \beta {\rm cos}\theta)^2}.
\end{equation}
The comoving synchrotron emissivity can be calculated using following formula,
\begin{eqnarray}
j_{\varepsilon^\prime}^{\rm syn} = \frac{\sqrt{3}}{4\pi} \frac{e^3 B}{m_e c^2 2 \pi \hbar \varepsilon^\prime} \int d\gamma_e n_{\gamma_e} G(x),
\end{eqnarray}
where
\begin{equation}
G(x)\approx\frac{1.81 e^{-x}}{(x^{-2/3} + (3.62/\pi)^2)^{1/2}},
\end{equation}
$x = \varepsilon^\prime / \varepsilon^\prime_c$ and  $\varepsilon^\prime_c = (3e\hbar B / 2 m_e c) \gamma_e^2 \beta^2$ is the critical energy~\citep{Rybicki:1986, Aharonian:2010va}.

The comoving SSC emissivity is estimated to be
\begin{eqnarray}
j_{\varepsilon^\prime}^{\rm SSC} = \frac{3}{16\pi} \sigma_T c \int d\gamma_e \frac{1}{\gamma_e^2} n_{\gamma_e} \int d\varepsilon_{\rm syn}^\prime \frac{\varepsilon^\prime}{\varepsilon_{\rm syn}^\prime}\frac{dn_{\rm syn}^\prime}{d\varepsilon_{\rm syn}^\prime} f(q, w),
\end{eqnarray}
where $\sigma_T$ is the Thomson cross section, $dn_{\rm syn}^\prime/d\varepsilon_{\rm syn}^\prime$ is the comoving synchrotron photon density, and
\begin{equation}
f(q, w) = 2q{\rm ln}q + (1+2q)(1-q) + \frac{1}{2} \frac{(wq)^2}{1+wq}(1-q),
\end{equation}
$q = \varepsilon^\prime / (4\gamma_e \varepsilon_{\rm syn}^\prime) (\gamma_e - \varepsilon^\prime/m_e c^2)$, and $w = 4\varepsilon_{\rm syn}^\prime \gamma_e /m_e c^2$~\citep{Blumenthal:1970gc}. 
Note the Thomson limit corresponds to $w \ll 1$, while the above expression is valid for all the value of $w$ as long as $\gamma_e \gg 1$~\citep{Blumenthal:1970gc}. 

The comoving EIC emissivity can be calculated as
\begin{eqnarray}\label{emissivity_EIC}
j_{\varepsilon^\prime}^{\rm EIC} &=& \frac{3}{8\pi} \sigma_T c (1-{\rm cos}~\theta_{\rm sc}^\prime) \int d\gamma_e n_{\gamma_e} \int dy A(y) \varepsilon_{\rm ph}^\prime \frac{dn_{\rm ph}^\prime}{d\varepsilon_{\rm ph}^\prime},
\end{eqnarray}
where $\varepsilon^\prime$ is the comoving EIC photon energy, $\varepsilon_{\rm ph}^\prime$ is the comoving seed photon energy, $dn_{\rm ph}^\prime/d{\varepsilon_{\rm ph}^\prime}$ is the comoving density of seed photons, $\theta_{\rm sc}^\prime$ is the scattering angle measured in the forward shock comoving frame relative to the direction of the photon beam, $y \equiv \xi m_e c^2/(2(1-{\rm cos}~\theta_{\rm sc}^\prime)\gamma_e \varepsilon_{\rm ph}^\prime (1-\xi))$,  $\xi \equiv \varepsilon^\prime /(\gamma_e m_e c^2)$, and $A(y) \equiv (1-\xi) [1-2y+2y^2+\xi^2/(2(1-\xi))]$~\citep{Aharonian:1981,Toma:2009mw, Murase:2010fq}.
Assuming the direction of the photon beam at the scattering point following the radial expansion of the jet, the observed flux can be evaluated at $\theta_{\rm sc}^\prime = \theta^\prime$ in the jet comoving frame where ${\rm cos}\theta_{\rm sc}^\prime = {\rm cos}\theta^\prime = (\mu - \beta)/(1-\beta \mu)$ with $\mu = {\rm cos}\theta$.

\begin{figure}
\includegraphics[width=0.49\textwidth]{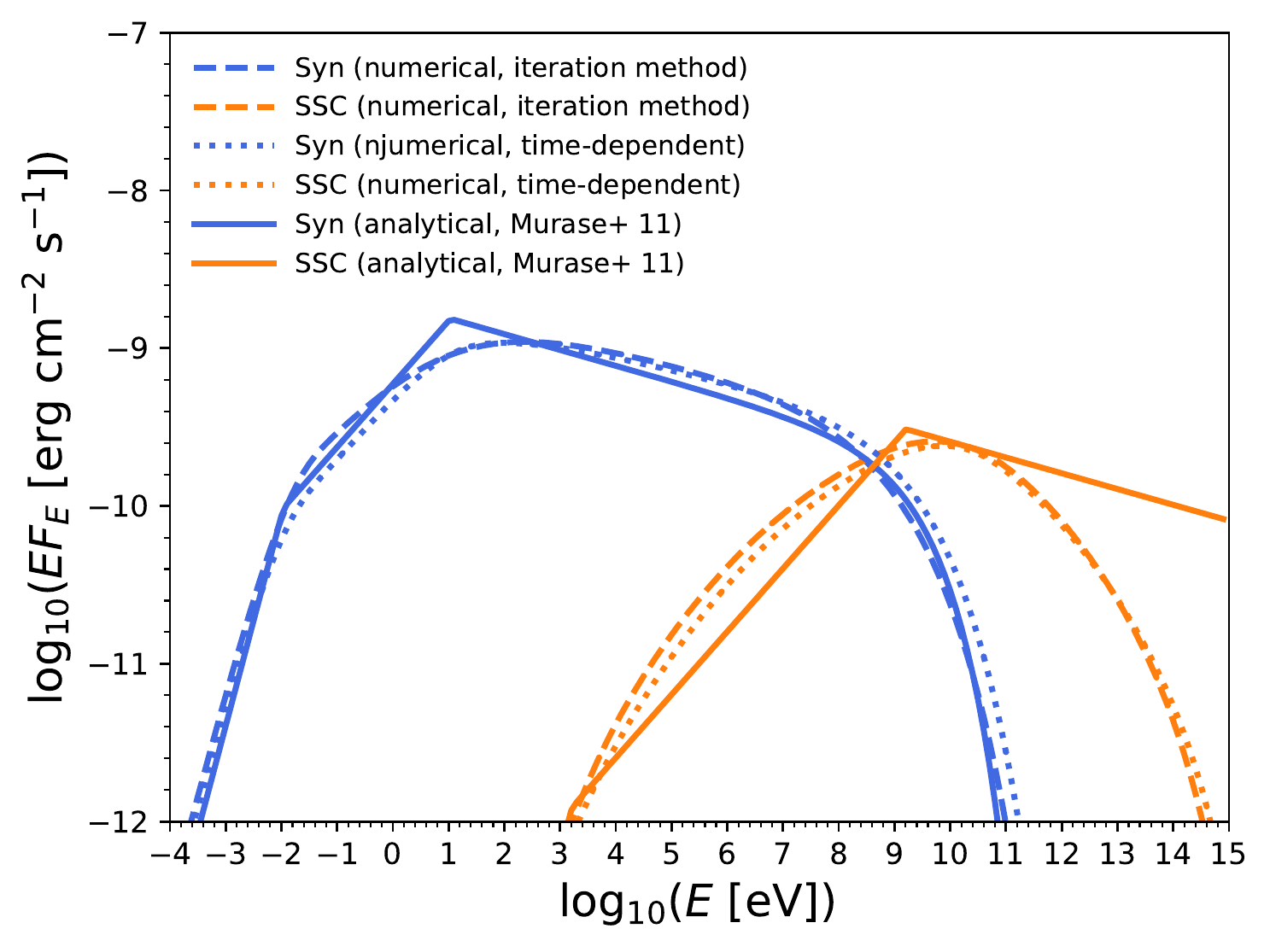}
\includegraphics[width=0.49\textwidth]{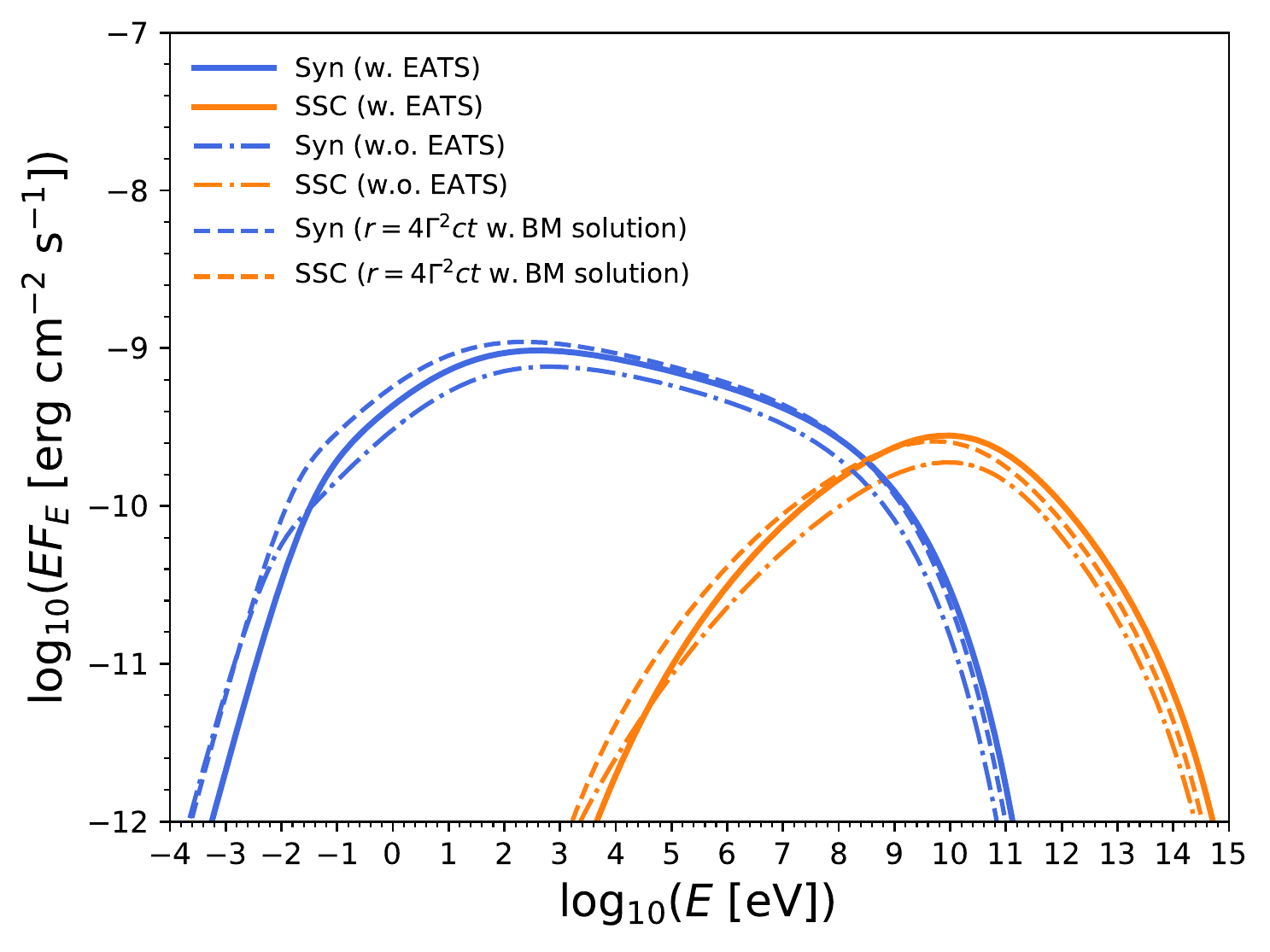}
\caption{\textit{Left pannel:} Comparison of energy spectra of synchrotron and SSC emission at $t = 10^4 \rm~s$ calculated in the single-zone model with both the numerical method (solid line) and the analytical method (dot-dashed line). The dotted line is calculated by the numerical method with a time-dependent electron energy spectrum via solving Equation~\ref{continuity-equation}. We also use the relation $r = 4\Gamma^2 c t$ in this panel. 
\textit{Right pannel:} Comparison of energy spectra of synchrotron and SSC emission at $t = 10^4 \rm~s$ calculated with EATS (solid line) and without EATS (dotted-dashed line) using the detailed dynamics described in Appendix~A. The dashed line is calculated by the single-zone model using the same dynamics as in left figure.
The relevant physical parameters are $\mathcal{E}_k = 1 \times 10^{53}\rm~erg$, $n_{\rm ex} = 1\rm~cm^{-3}$, $\epsilon_e = 0.05$, $f_e = 1$, $\epsilon_B = 1 \times 10^{-2}$, $s=2.2$, $\Gamma_0 = 100$, and $z = 0.0785$. 
\label{fig:spectrum_compare}}
\end{figure}

In Figure~C\ref{fig:spectrum_compare} left, we compare results of different methods for the single-zone model. The iteration method used in this work agrees with the method of solving the time-dependent equation. Both of the numerical results agree with the analytical formula. The analytical synchrtron and SSC spectra are taken from \citet{Murase:2010fq}, where the BM solution with $r = 4\Gamma^2 c t$ is used for dynamics~\citep{Blandford:1976uq, Sari:1997qe, Waxman:1997yv} and the synchrotron peak flux is evaluated with a correction factor introduced in \citet{Wijers:1998st}. Note that the SSC spectrum in the Thomson limit is shown here, although \citet{Murase:2010fq} also considered spectral suppression due to the Klein-Nishina effect analytically. 
In Figure~C\ref{fig:spectrum_compare} right, we compare results of the single-zone model with $r=4\Gamma^2 c t$ to those of the detailed afterglow dynamics moel with/without EATS. This panel demonstrates that the detailed model with EATS agrees with the standard single-zone afterglow model in the relativistic limit, although our detailed method has an advantage that it can be used both relativistic and non-relativistic regimes consistently. Note that we our synchrotron energy spectra calculated with EATS are also consistent with the results of \cite{Granot:2001ge}.

Photons with energy beyond the pair production threshold will undergo internal absorption by ambient photons inside the source and external absorption by extragalactic background light (EBL) during their propagation to Earth. The threshold energy can be estimated from the kinematic condition, $\varepsilon^\prime \varepsilon_{\rm seed}^\prime \gtrsim (m_e c^2)^2$, where $\varepsilon^\prime$ is the comoving high-energy photon energy and $\varepsilon_{\rm seed}^\prime$ is the comoving target photon energy. As in \cite{Murase:2010fq}, we use the $\gamma\gamma$ optical depth,
\begin{eqnarray}
\tau_{\gamma \gamma} = \frac{\tilde{\Delta}}{2} \int_{-1}^1 d\mu (1 -\mu) \int d\varepsilon_{\rm seed}^\prime \frac{dn}{d\varepsilon_{\rm seed}^\prime} \sigma_{\gamma \gamma} (S),
\end{eqnarray}
where
\begin{equation}
\sigma_{\gamma \gamma} (S) = \frac{3}{16} \sigma_T (1-\beta_{\rm cm}^2) [(3-\beta_{\rm cm}^4) {\rm ln}\left( \frac{1 + \beta_{\rm cm}}{1 - \beta_{\rm cm}}\right) - 2\beta_{\rm cm} (2-\beta_{\rm cm}^2)]
\end{equation}
is the two-photon annihilation cross section, $\beta_{\rm cm} = \sqrt{1-S^{-1}}$, and $S = (1/2)\varepsilon_{\rm seed}^\prime \varepsilon^\prime (1-\mu)$ is the Mandelstam variable. 
Here, $\tilde{\Delta}$ is the comoving width, and the attenuation at a given EATS includes from contributions with different values of $\tilde{\Delta}$. For simplicity, we adopt $\tilde{\Delta}\approx ct/(1+z)$ and ignore effects of electromagnetic cascades.
The photon spectrum after the internal $\gamma \gamma$ absorption can be estimated as $F_\varepsilon^{\gamma \gamma} = F_\varepsilon / (1 + \tau_{\gamma \gamma})$.
The EBL is mainly composed of infrared and optical photons, and we use the low-IR model calculated in \cite{Kneiske:2003tx}. The observed energy spectrum is
\begin{eqnarray}
F_E^{\rm ob} = F_E^{\gamma \gamma} e^{-\tau_{\gamma\gamma}^{\rm EBL}(E, z)},
\end{eqnarray}
where $\tau_{\gamma\gamma}^{\rm EBL}(E, z)$ is the optical depth at given observed photon energy $E=\varepsilon/(1+z)$ and redshift $z$.

\end{document}